\documentclass[%
superscriptaddress,
showpacs,preprintnumbers,
amsmath,amssymb,
aps,
prl,
linenumbers,
twocolumn
]{revtex4}
\usepackage{color}
\usepackage{graphicx}
\usepackage{dcolumn}
\usepackage{bm}

\begin {document}

\title{Origin of subdiffusion of water molecules on cell membrane surfaces}

\author{Eiji Yamamoto}
\author{Takuma Akimoto}
\affiliation{%
  Department of Mechanical Engineering, Keio University, Yokohama, Japan
}%

\author{Masato Yasui}
\affiliation{%
  Department of Pharmacology, School of Medicine, Keio University, Shinjuku-ku, Tokyo, Japan
}%

\author{Kenji Yasuoka}
\affiliation{%
  Department of Mechanical Engineering, Keio University, Yokohama, Japan
}%



\begin{abstract}
Water molecules play an important role in providing unique environments for biological reactions on cell membranes. 
It is widely believed that water molecules form bridges that connect lipid molecules and stabilize cell membranes. 
Using all-atom molecular dynamics simulations, we show that translational and rotational diffusion of water molecules on lipid membrane surfaces exhibit subdiffusion.
Moreover, we provide evidence that both divergent mean trapping time (continuous-time random walk) and long-correlated noise (fractional Brownian motion) contribute to this subdiffusion.
These results suggest that subdiffusion on cell membranes causes the water retardation, an enhancement of cell membrane stability, and a higher reaction efficiency.
\end{abstract}

\maketitle


Water molecules around cell membranes are important for stability and dynamics of self-assembled lipid structures.
Such water molecules form a bridge network that connects lipid molecules~\cite{Pasenkiewicz-GierulaTakaokaMiyagawaKitamuraKusumi1997}.
Water molecules form local hydration structures depending on the lipid head groups~\cite{DamodaranMerz1993,AlperBassolino-KlimasStouch1993,LopezNielsenKleinMoore2004} and are weakly aligned by charges on the lipid head group~\cite{NagataMukamel2010,MondalNihonyanagiYamaguchiTahara2010,ChenHuaHuangAllen2010,MondalNihonyanagiYamaguchiTahara2012}.
Thus, it is difficult for water molecules on the membrane surface to diffuse freely on the surface of membranes.
Qualitatively, translational and rotational motions of water molecules near membranes are slower than those in the bulk~\cite{R'ogMurzynPasenkiewicz-Gierula2002,BhideBerkowitz2005,murzyn2006dynamics,YamamotoAkimotoHiranoYasuiYasuoka2013}.
Although static properties of such water molecules have been known from experiments, little is known about how water molecules actually diffuse on the membrane surface.

In usual case, diffusion can be characterized by the ensemble-averaged mean square displacement (MSD), {\it i.e.}, $\langle \mbox{\boldmath $r$}^2(t) \rangle =2Dt$, where $D$ is the diffusion constant. However, extensive experimental studies show subdiffusion,  
\begin{equation}
\langle \mbox{\boldmath $r$}^2(t) \rangle \simeq K_\alpha t^\alpha \quad {\rm with} \quad  0<\alpha<1,
\end{equation}
where $\alpha$ is the subdiffusive exponent and $K_\alpha$ the generalized diffusion constant.
There are three well-known stochastic models  of subdiffusions with different mechanisms: fractional Brownian motion (FBM)~\cite{Kolmogorov1940,MandelbrotVan1968}, diffusion on a fractal lattice~\cite{Ben-AvrahamHavlin2000}, and continuous-time random walk (CTRW)~\cite{MetzlerKlafter2000}.
Because these models have different physical nature, revealing the origin is significant to understand physical properties~\cite{MasonWeitz1995,TejedorB'enichouVoituriezJungmannSimmelSelhuber-UnkelOddershedeMetzler2010}.
In particular, the physical origin of subdiffusion in living cells has been extensively studied~\cite{GoldingCox2006,JeonTejedorBurovBarkaiSelhuber-UnkelBerg-SOrensenOddershedeMetzler2011,WeigelSimonTamkunKrapf2011,BarkaiGariniMetzler2012,TabeiBurovKimKuznetsovHuynhJurellerPhilipsonDinnerScherer2013}.
Previously, subdiffusive motion of water molecules on the surface of a membrane were reported~\cite{YamamotoAkimotoHiranoYasuiYasuoka2013,HansenGekleNetz2013}.
However, the origin of this water subdiffusion remains unclear.

In general, it is difficult to identify the mechanism underlying subdiffusion.
Ergodic and aging properties play an important role in clarifying the physical origin.
It is known that FBM motion is ergodic, whereas under confinement a power-law relaxation of time-averaged mean square displacement occurs for a model related to FBM~\cite{JeonMetzler2012,JeonLeijnseOddershedeMetzler2013}.
Since ordinary ergodicity, where the time averages are equal to the ensemble average, holds for FBM, and diffusion on a fractal lattice~\cite{DengBarkai2009,JeonMetzler2010}, the dominant feature of CTRW with a divergent mean trapping time is aging and weak ergodicity breaking~\cite{Bouchaud1992,HeBurovMetzlerBarkai2008,LubelskiSokolovKlafter2008,MiyaguchiAkimoto2013}.
Such phenomena are also observed in a range of stochastic models different from the CTRW such as random walk with static disorder~\cite{MiyaguchiAkimoto2011a}, random walks with correlated waiting times~\cite{TejedorMetzler2010,MagdziarzMetzlerSzczotkaZebrowski2012}, spatially correlated random walks~\cite{CherstvyChechkinMetzler2013}, aging walks~\cite{LomholtLizanaMetzlerAmbjornsson2013}, and stored-energy-driven L\'evy flight~\cite{AkimotoMiyaguchi2013}.

Divergence of the mean trapping time is attributed to a power law in the trapping-time distribution.
Power laws are often observed in biological phenomena~\cite{WongGardelReichmanWeeksValentineBauschWeitz2004,HijkoopDammersMalekCoppens2007,WeigelSimonTamkunKrapf2011,AkimotoYamamotoYasuokaHiranoYasui2011}.
One of the mechanisms generating a power-law trapping-time distribution is a random-energy landscape~\cite{Bouchaud1992}. 
There are many binding sites in one-dimensional diffusion along DNA, two-dimensional diffusion on the plasma membrane, and three-dimensional diffusion in the cytoplasm~\cite{Saxton2007}.
If the potential depth of each binding site is randomly distributed according to the exponential distribution, the distribution of the trapping times for which particles are trapped in the binding sites follows a power law~\cite{BardouBouchaudAspectTannoudji2002}.

In this letter, we perform molecular dynamics (MD) simulations on two systems of water molecules plus membranes, of either palmitoyl-oleoyl-phosphocholine (POPC) or palmitoyl-oleoyl-phosphatidylethanolamine (POPE), at the temperature 310~K to investigate the diffusion of water molecules on the membrane surface~(Fig.~1A).
Here, we report on subdiffusion of water molecules on the membrane surface.
Furthermore, we show that the subdiffusion is attributed to the divergent mean trapping time and anti-correlated noise, {\it i.e.}, a combination of CTRW and FBM scenarios.
We confirm there are no qualitative differences about subdiffusive behavior despite of the different water structure at ammonium head groups of POPC and POPE~\cite{DamodaranMerz1993}.
 
\begin{figure}[tb]
\begin{center}
\includegraphics[width=85 mm,bb= 0 0 390 315]{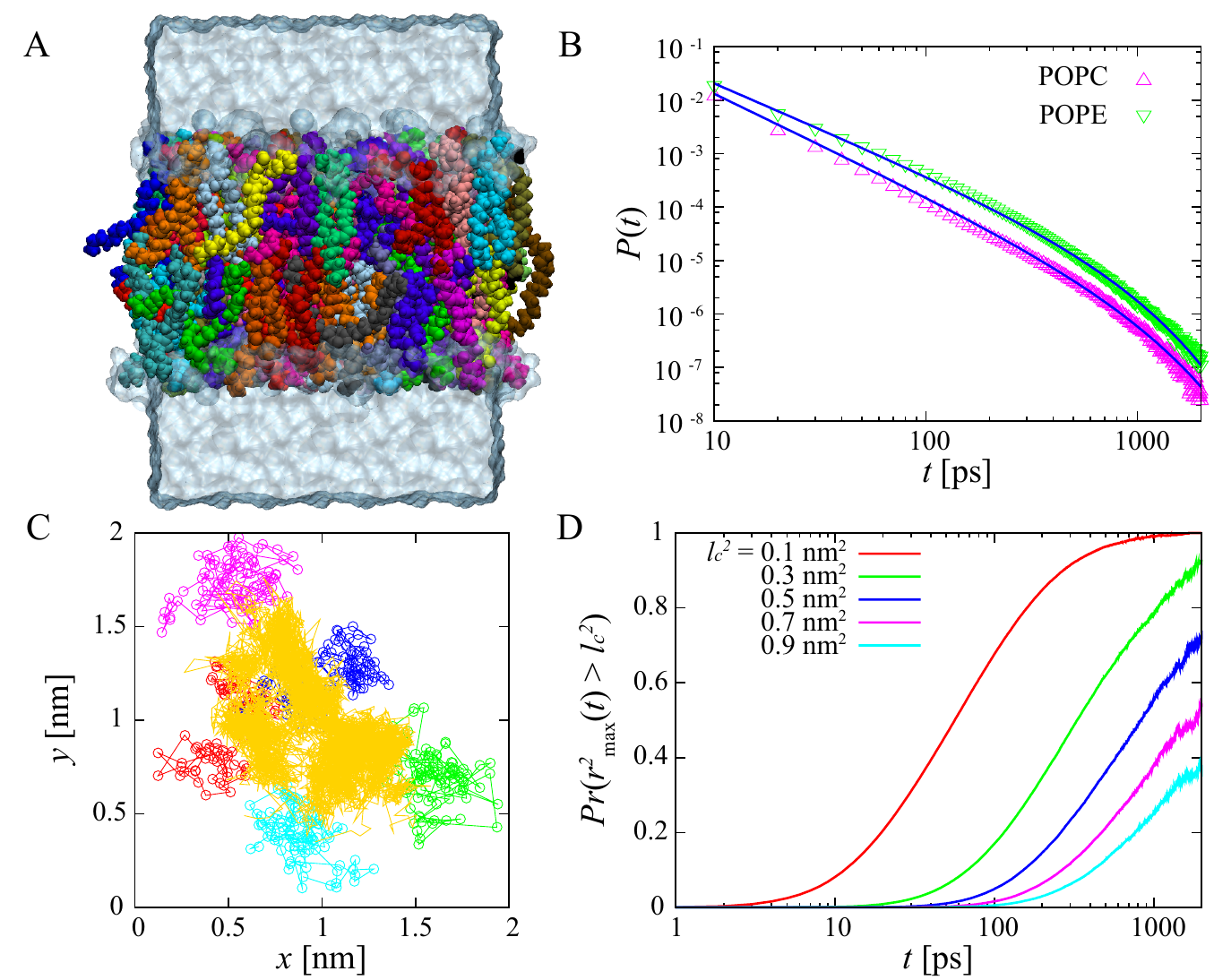}
\caption{Diffusion of water molecules on lipid membrane surfaces.
(A)~Configuration of POPC bilayer.
Each color represents a different phospholipid.
Explicit water molecules correspond to the upper and lower transparent coatings.
(B)~Residence time PDFs $P(t)$ of water molecules on the membrane surfaces.
Solid lines are fitting curves by power-law distributions with exponential cutoffs: $P(t)=At^{- \beta} \exp (-Bt)$ (POPC: $\beta = 1.9$, $B=0.0013$, POPE: $\beta = 1.7$, $B=0.0016$).
(C)~Lateral trajectory of a water molecule (yellow) tracked for 9~ns on a POPE membrane surface.
Circles with lines represent trajectories of the C2 atom (see Fig.~S1) in different lipid molecules.
(D)~Fraction of water molecules traversing a certain distance $l_c$.
Each color represents different $l_c$ values (see key legend).}
\vspace{-7mm}
\end{center}
\end{figure}

\begin{figure}[tb]
\begin{center}
\includegraphics[width=85 mm,bb= 0 0 392 310]{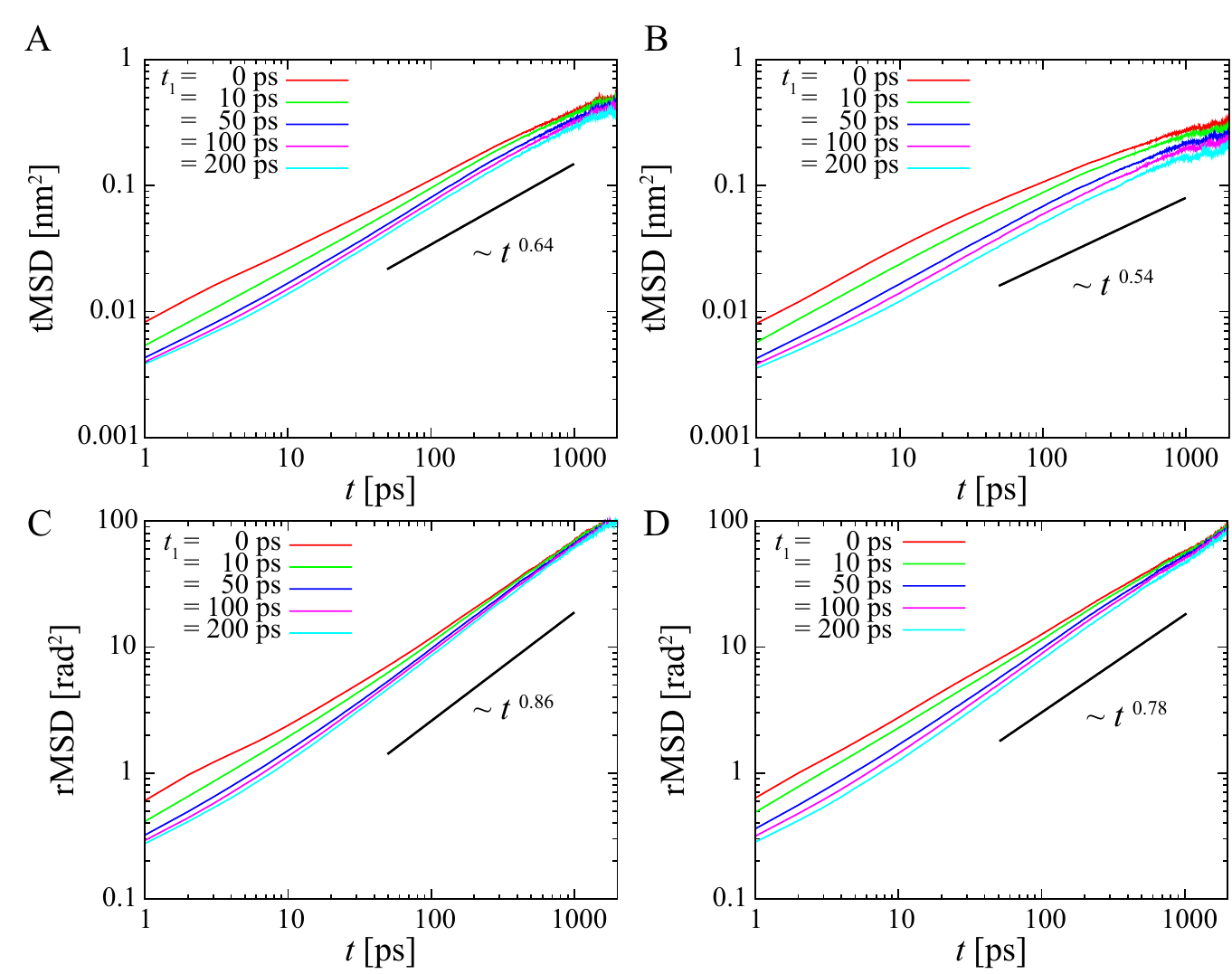}
\caption{(A)~Ensemble-averaged tMSD and (C)~rMSD of water molecules on a POPC membrane surface.
(B) and (D) are the tMSD and rMSD on a POPE membrane surface.
The slope of the solid lines are fitted in the time interval from 50 to 1000~ps for $t_1 = 200$~ps.
The different colored lines correspond to different measurement starting times $t_1$.}
\vspace{-7mm}
\end{center}
\end{figure}

{\it Diffusions of Water Molecules on Membrane Surfaces.$-$}Water molecules forming the bridges connecting lipid molecules on the membrane surface do not diffuse.
This bridge is formed by hydrogen bonds between the water molecules and head groups of the lipid molecules. These hydrogen bond interactions create a complicated and random potential surface over the membrane.
To investigate the diffusivity of water molecules on the membrane surface, we define surface water molecules as water molecules for which the oxygens remain continuously within interatomic distances of 0.35~nm from atoms (oxygen, phosphorus, nitrogen, and carbon atoms) in the lipid molecules.
In what follows, we use trajectories of the water and lipid molecules where the position of the center of mass of the membrane is subtracted.

First, we consider the residence time distribution of water molecules on the membrane surface, where the residence time is defined as the duration for which a water molecule remains on the membrane surface.
As shown in Fig.~1B, the probability density functions (PDFs) of the residence times on the POPC and POPE bilayers follow power-law distributions with exponential cutoffs in their tails.
Mean residence times of water molecules on POPC and POPE bilayers are 7.0 and 9.3~ps, respectively.
However, some water molecules reside on the membrane surfaces for more than 1~ns.
Figure~1C shows a lateral trajectory of a water molecule trapped on the POPE bilayer surface for 9~ns.
Surprisingly, water molecules do diffuse widely on the membrane surface while trapped on it.
In other words, a water bridge connecting lipid molecules in a membrane is not static but dynamical.
Indeed, diffusion distances on the membrane surface lengthen with increasing residence times.
Figure~1D shows the probability that the maximal excursion distance for water molecules is greater than $l_c$, $P(r^2_{\rm max}(t) > 2l^2_c)$, where the maximal excursion distance is defined by $r_{\rm max}(t) = {\rm max} \{ r(t') : 0 \leq t' \leq t \}$ with $r(t) = \sqrt[]{x(t)^2 + y(t)^2}$.
About 40~\% of all water molecules can diffuse above 0.5~${\rm nm}^2$ at 600~ps even if the water molecules remain on the membrane surface.
This implies that water molecules can diffuse beyond a lipid molecule in the membranes, because the area per lipid is about 0.5-0.7~${\rm nm}^2$.
Although many water molecules diffuse within the area of a lipid by forming a bridge and a hydration shell~\cite{DamodaranMerz1993,AlperBassolino-KlimasStouch1993}, some water molecules diffuse by interchanging the water bridge while remaining on the membrane surface.
Thus, we found a water-bridge interchange dynamics for the first time.


{\it Translational and Rotational Subdiffusion of Water Molecules.$-$}To investigate the diffusion of water molecules on the membrane surfaces, we consider translational as well as rotational diffusions of the water molecules.
The ensemble-averaged lateral translational MSD (tMSD) is defined as 
\begin{equation}
\left< l^2(t) \right> = \frac{1}{2} \langle \left\{ x(t+t_0) - x(t_0) \right\} ^2 + \left\{ y(t+t_0) - y(t_0) \right\}^2 \rangle,
\end{equation}
where $t_0$ is the time when water molecules enter the membrane surfaces and $\langle \ldots \rangle$ is the average with respect to captured and reflected water molecules impinging on the membrane surface.
If exiting from the membrane surfaces, water molecules are excluded from the ensemble.
In considering rotational diffusion, we define $\delta \theta (t) \equiv \cos^{-1} \left( \overrightarrow{\mu}(t) \cdot \overrightarrow{\mu}(t + \delta t) \right)$ and direction $\overrightarrow{p} (t) \equiv \overrightarrow{\mu}(t) \times \overrightarrow{\mu}(t+\delta t)$, where $\overrightarrow{\mu}(t)$ is the dipole vector of a water molecule at time $t$.
The vector $\overrightarrow{\varphi}(t)\equiv \int^{t_0+t}_{t_0} \delta \theta(t') \overrightarrow{p}(t')dt'$ gives us the trajectory representing the rotational motion.
Then, the ensemble-averaged rotational mean-squared displacement (rMSD)~\cite{MazzaGiovambattistaStarrStanley2006} is given by
\begin{equation}
\left< \varphi^2(t) \right> = \left\langle \left| \overrightarrow{\varphi} (t) - \overrightarrow{\varphi} (0) \right| ^2 \right\rangle.
\label{eq:ro_MSD}
\end{equation}

In CTRW, the MSD is suppressed with increase of the starting time $t_1$ of a measurement~\cite{Barkai2003}.
This behavior is called {\it aging}.
To investigate aging, we consider the dependence of the MSDs on the starting time of a measurement.
Here, we consider ${\rm tMSD}(t;t_1) = \langle \{ x(t+t_0+t_1) - x(t_0+t_1) \} ^2 + \{ y(t+t_0+t_1) - y(t_0+t_1) \} ^2 \rangle /2$ and ${\rm rMSD}(t;t_1) = \left\langle | \overrightarrow{\varphi} (t + t_0 + t_1) - \overrightarrow{\varphi} (t_0 + t_1) | ^2 \right\rangle $, where $t_1$ corresponds to times after entering the membrane surface at $t_0$. 
Figure~2 shows the MSDs measured after time $t_1$ from $0$ to $200$~ps. 
Translational motions of water molecules exhibit subdiffusion as in diffusion of lipid molecules~~\cite{WeissHashimotoNilsson2003,FlennerDasRheinstadterKosztin2009,AkimotoYamamotoYasuokaHiranoYasui2011,KnellerBaczynskiPasenkiewicz-Gierula2011,JeonMonneJavanainenMetzler2012}.
Whereas the subdiffusive exponents in the tMSDs decrease as time $t$ increases, the rMSDs show subdiffusion with a constant subdiffusive exponent.
For tMSD and rMSD, water molecules on POPC bilayers are faster than those on POPE bilayers.
This is because hydrogen bonds between choline groups and water molecules in POPC bilayers are weaker than those in POPE bilayers because methyl groups are present in the choline group of POPC.
Moreover, as seen in Fig.~2, both tMSD and rMSD depend on the starting time of a measurement $t_1$. 
Both MSDs become smaller the later $t_1$ becomes.
For $t_1 > 50$~ps, unlike CTRW, MSDs do not strongly depend on $t_1$.
Therefore, this aging will be affected by a non-equilibrium initial condition when water molecules attach to the membrane surfaces.
We note that MSDs calculated after equilibration on membrane surfaces also decrease according to $t_1$ (see Fig.~S2).

\begin{figure}[bt]
\begin{center}
\includegraphics[width=85 mm,bb= 0 0 318 314]{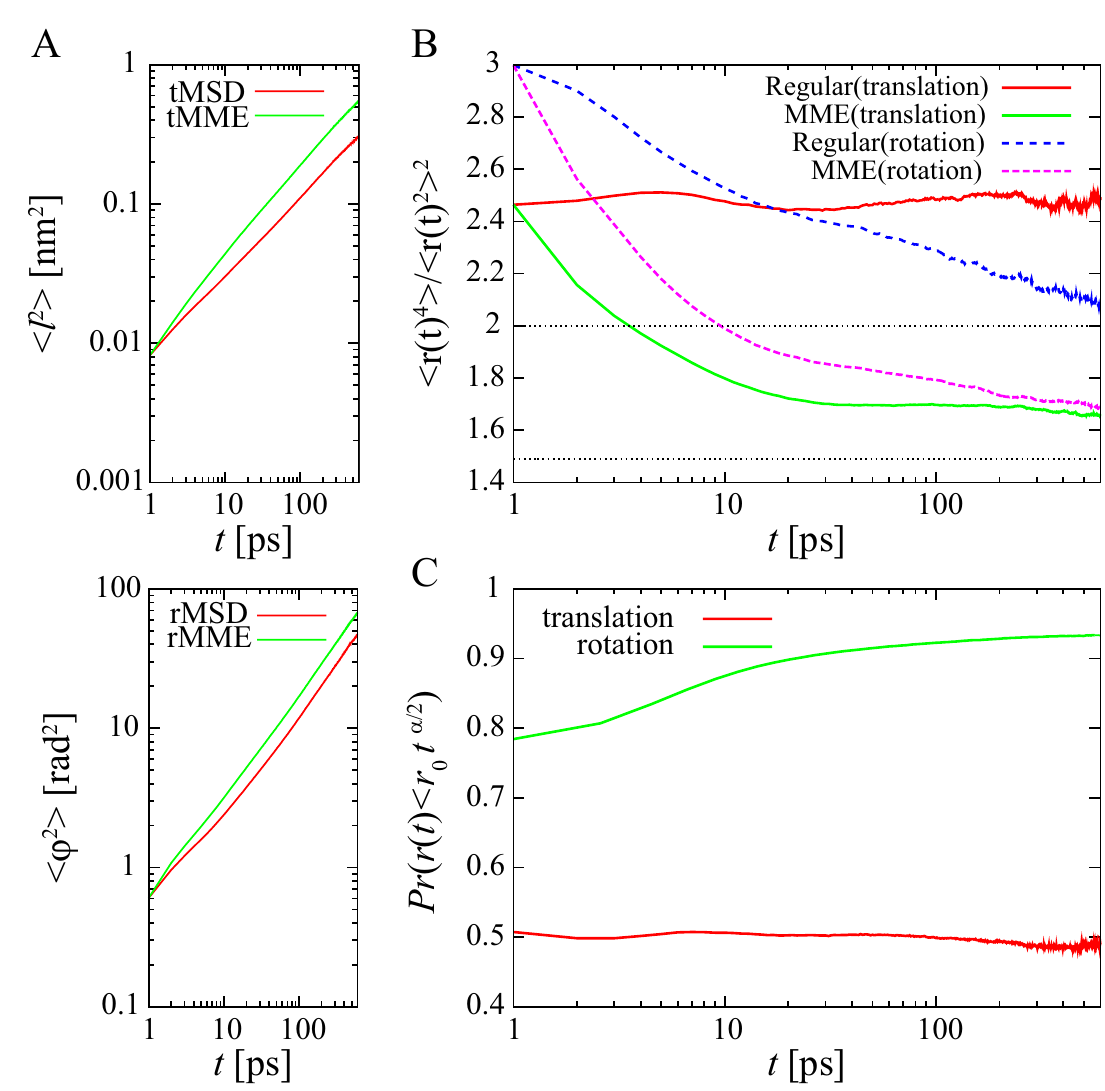}
\caption{Quantitative analysis of trajectories of water molecules on the POPC membrane surface. 
(A)~MSD and second MME moment as functions of time $t$ for translational and rotational diffusions.
(B)~Regular and MME moment ratios for translational and rotational motions.
Horizontal lines are ratios 2 and 1.49.
(C)~Probability of water molecules to be in a sphere of growing radius $r_0 t^{\alpha/2}$.
The value of $\alpha$ is based on fitted values 0.56 and 0.76 for translational and rotational motions in the time interval from 10 to 1000~ps for $t_1 = 0$~ps, respectively.}
\vspace{-7mm}
\end{center}
\end{figure}

{\it Origin of Subdiffusive Motion of Water Molecules.$-$}To clarify the origin of subdiffusive motions of water molecules on membrane surfaces, we perform a mean maximal excursion (MME) analysis~\cite{TejedorB'enichouVoituriezJungmannSimmelSelhuber-UnkelOddershedeMetzler2010}.
The MME analysis provides us an information on the physical nature of the underlying subdiffusive processes by using trajectories only.
In Fig.~3A, the translational and rotational MSDs, $\left< l^2(t) \right>$ and $\left< \varphi^2(t) \right>$, and the MME second moments, $\left< l^2(t)_{\rm max} \right>$ and $\left< \varphi^2(t)_{\rm max} \right>$, grow sublinearly with time, where $\left< l^2(t)_{\rm max} \right>$ and $\left< \varphi^2(t)_{\rm max} \right>$ are the ensemble averages of $l_{\rm max}(t) = {\rm max} \{ l(t') : 0 \leq t' \leq t \}$ and $\varphi_{\rm max}(t) = {\rm max} \{ \varphi(t') : 0 \leq t' \leq t \}$, respectively.
For about $t>30$~ps, the subdiffusive exponents of MSDs are almost the same as those of the MME second moment.
This result suggests that a fractal or CTRW feature appears over relatively large-time intervals.
Moreover, Fig.~3B shows that the regular moment ratios $\left< l^4(t) \right> / \left< l^2(t) \right>^2$ and $\left< \varphi^4(t) \right> / \left< \varphi^2(t) \right>^2$ fluctuate above 2 and that the MME moment ratios $\left< l^4(t)_{\rm max} \right> / \left< l^2(t)_{\rm max} \right>^2$ and $\left< \varphi^4(t)_{\rm max} \right> / \left< \varphi^2(t)_{\rm max} \right>^2$ fluctuate above 1.49.
This result suggests CTRW scenario and excludes FBM and fractal scenarios.
Figure~3C shows that the probability for water molecules to be in a sphere of growing radius $r_0 t^{\alpha/2}$ is almost constant over $t$, while for rotational diffusions, the probability below 20~ps increases because of a change in the subdiffusive exponent.
This result suggests CTRW or FBM scenarios and excludes fractal scenario.
The above results are summarized in Table 1 in supporting information.
These results strongly support the CTRW scenario for large-time intervals.

To validate the CTRW scenario, we consider the time-averaged mean square displacements (TAMSDs) defined by $\overline{\delta^2(\Delta ; t)} = \left( \overline{\delta_x^2(\Delta ; t)} + \overline{\delta_y^2(\Delta ; t)} \right) / 2$ and $\overline{\delta_{\varphi}^2(\Delta ; t)} = \frac{1}{t - \Delta } \int_0^{t - \Delta } | \vec{\varphi}(t' + \Delta )- \vec{\varphi}(t') |^2dt'$ for translational and rotational motions, respectively, where $t$ is the measurement time and $\overline{\delta_x^2(\Delta ; t)} = \int_0^{t - \Delta } \{ x(t' + \Delta )-x(t')\}^2dt'/(t-\Delta)$.
TAMSDs for trajectories of water molecules residing on the surface of the membrane longer than 2000~ps for both translational and rotational motions are shown in Figs.~4A and 4B, respectively. 
Unlike CTRW, where the TAMSD grows linearly with $\Delta$, TAMSDs do not show a linear scaling over short-time durations.
Because the TAMSD shows subdiffusion in FBM, i.e., sublinear scaling of $\Delta$, translational and rotational motions have a FBM characteristic over short-time durations of $\Delta$.
However, rotational TAMSDs show normal diffusion (linear scaling of $\Delta$) as expected by CTRW, whereas translational TAMSDs do not show normal diffusion. 
The mean rotational TAMSDs crossover from sublinear to linear (see Fig.~S4).
The crossover points at around 10~ps are coincident with the relaxation time for the orientational correlation functions of water molecules on the membrane surfaces~\cite{YamamotoAkimotoHiranoYasuiYasuoka2013}.
Because the sublinear growth of the TAMSDs suggests FBM, the dynamics of water molecules will be affected by viscoelasticity.

Figures~4C and 4D show the aging plots for translational and rotational TAMSDs on the POPC membrane surface, i.e., the ensemble average of the TAMSD as a function of the measurement time $t$, for different measurement starting times $t_1$.
Whereas the ensemble averages of translational and rotational TAMSDs show power-law decays: $\langle \overline{\delta^2 (\Delta;t)} \rangle \propto t^{- \gamma_1}$ and $\langle \overline{\delta_{\varphi}^2 (\Delta;t)} \rangle \propto t^{-\gamma_2}$ for $t_1 < 50$~ps, those do not decay for $t_1 > 50$~ps.
In CTRW, the ensemble average of a TAMSD decays as $\langle \overline{\delta^2 (\Delta;t)} \rangle \propto t^{-(1-\alpha)}$~\cite{HeBurovMetzlerBarkai2008}, where $\alpha$ is the power-law exponent for the trapping-time PDF.
However, recently, it is shown that CTRW with strong noisy fluctuations do not show the aging of TAMSD, whereas MSD still shows aging~\cite{JeonBarkaiMetzler2013}.
Thus, the power-law decays of ensemble average of TAMSDs  for $t_1<50$~ps are attributed to non-equilibrium initial conditions of water molecules on the membrane surfaces.
This is because mean velocity of bulk water molecules is higher than those on the membrane surfaces.
We note that MSDs show aging in our simulations even when an initial non-equilibrium state is skipped (see Fig.~S2).

Together with the MME analysis, it is physically reasonable to consider that the origin of the observed subdiffusion is a combination of CTRW and FBM.
Although we do not provide a direct evidence of aging effect, results in noisy CTRW~\cite{JeonBarkaiMetzler2013} assist a suggestion that aging due to CTRW is inherent in water dynamics on the membrane surfaces.
We note that non-equilibrium conditions of water molecules on the membrane surface are compatible with an equilibration of the total system.
As shown in the supporting information~(Fig.~S8, S9), total systems are equilibrated whereas TAMSDs show apparent aging (see aging plot in Fig.~4).
This apparent inconsistency can be resolved by dissociation of water molecules from the membrane surfaces.
In fact, because water molecules can dissociate from the membrane surfaces and the mean residence time is finite, the system can be equilibrated.

The distribution of waiting times contributes to CTRW arising from random binding and unbinding of water molecules from the lipid surface.
Moreover, translational motions of water molecules forming the water bridge are affected by lipid motions in lipid membranes which are governed by FBM motions~\cite{AkimotoYamamotoYasuokaHiranoYasui2011,KnellerBaczynskiPasenkiewicz-Gierula2011,JeonMonneJavanainenMetzler2012}.
Unlike CTRW, where a trapped state simply does not move, it is difficult to estimate exact trapping times in such situations.
Thus, we do not observe power-law trapping-time distributions in lateral motions (see Fig.~S6) because water molecules on the membrane surface can move during a trapped state.
We also confirmed that there are no significant differences in the water behavior on both POPC and POPE lipid membranes (see Fig.~S2-S7).

\begin{figure}[tb]
\begin{center}
\includegraphics[width=87 mm,bb= 0 0 428 277]{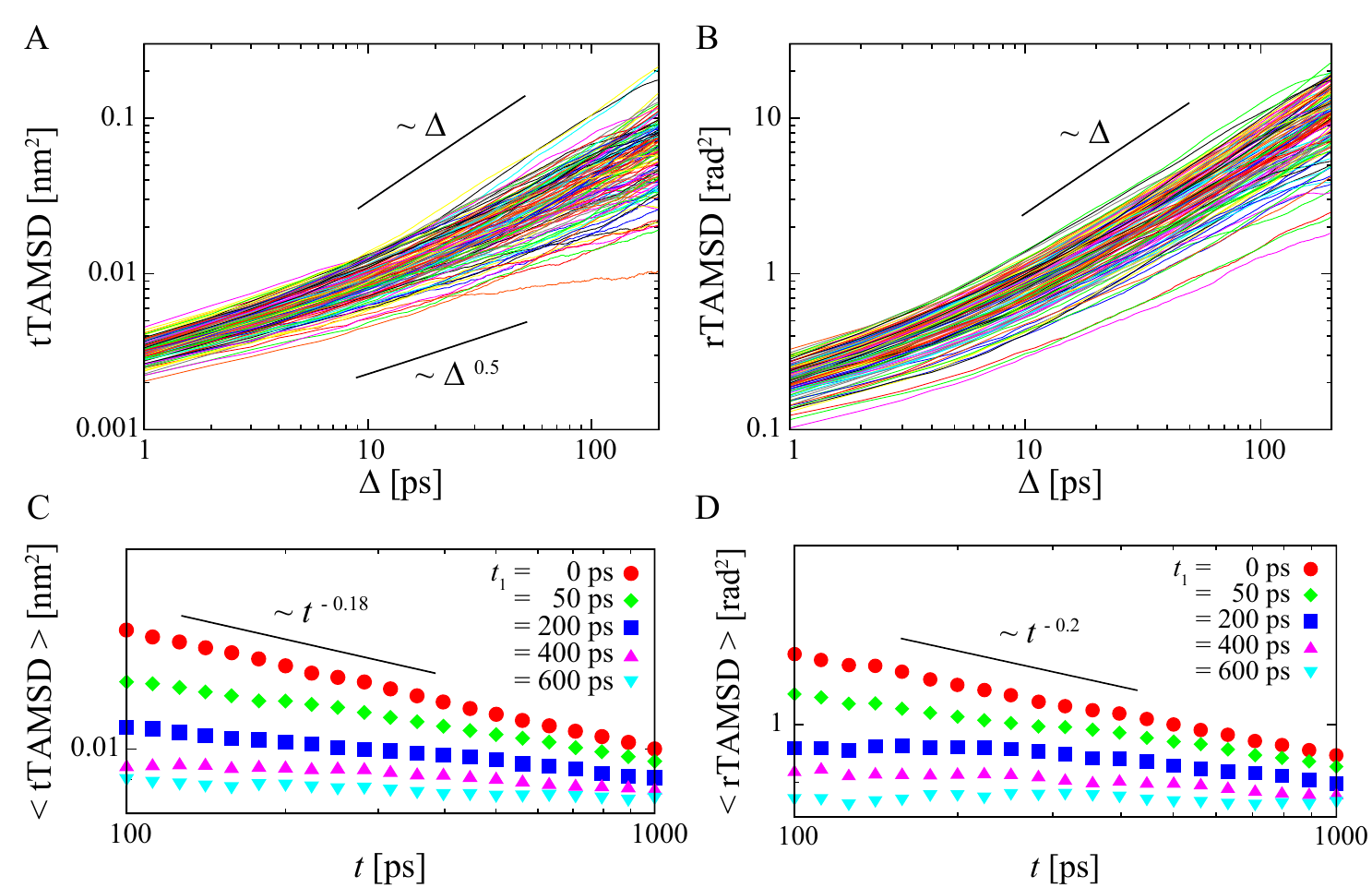}
\caption{(A)~Translational and (B)~rotational TAMSDs of water molecules on the POPC membrane surface.
The different colored lines show 128 trajectories of water molecules.
(C)~Aging plot for translational and (D)~rotational TAMSD for $\Delta = 10$~ps.
The different colored symbols correspond to different measurement starting times $t_1$.
For reference, the power-law decays are represented by solid lines.}
\vspace{-7mm}
\end{center}
\end{figure}


In summary, we have shown that water molecules on membrane surfaces can diffuse laterally while connected as part of a bridging network to lipid molecules in membrane.
This interchanging dynamics in the water bridge network can be described by CTRW.
Furthermore, we have found translational and rotational subdiffusion of water molecules on the membrane surfaces. 
These subdiffusions originate from a combination of CTRW and FBM, which are attributed to long-time trapping by the membrane surface and viscoelasticity of lipid bilayers, respectively.
Such a subdiffusive process has been observed in experiments of intracellular transport of insulin granules~\cite{TabeiBurovKimKuznetsovHuynhJurellerPhilipsonDinnerScherer2013}.

What is a biological significance of anomalous diffusion of water molecules on cell membrane surfaces?
To recognize a target, biomolecules diffuse slowly around the target, and may be guided by the behavior of water molecules in the target vicinity.
For example, water retardation around a metalloenzyme active site assists enzyme-substrate interactions~\cite{GrossmanBornHeydenTworowskiFieldsSagiHavenith2011}.
In a stochastic model, the probability of finding a nearby target is explicitly increased by subdiffusion~\cite{GuigasWeiss2008}.
Biological reactions such as ligand-receptor interactions and enzymatic reactions occur on cell membranes and depend upon encounters between biomolecules.
The roles of the surrounding water molecules depend upon the structure and dynamics of water molecules in the hydration layer of the membranes.
Anomalous diffusion of water molecules on the membrane surfaces increase the chance to bind the membrane surfaces and cause water retardation.
As a result, water molecules form bridges that connect lipid molecules and stabilize cell membranes. 
Moreover, the water retardation contributes to higher efficiency of biological reactions on cell membranes.

This work is supported by the Core Research for the Evolution Science and Technology (CREST) of the Japan Science, Technology Corporation (JST), and Keio University Program for the Advancement of Next Generation Research Projects.


\end{document}